\documentclass[a4paper,fleqn,usenatbib]{mnras}

\usepackage[T1]{fontenc}
\usepackage{ae,aecompl}

\usepackage{graphicx}	
\usepackage{amsmath}	
\usepackage{amssymb}	

\title[Magnetar outburst and spin-down glitch]{Magnetar outburst and spin-down glitch}

\author[Tong \& Huang]{H. Tong$^{1}$\thanks{E-mail: htong\_2005@163.com}, L. Huang$^{2,3,4}$
\\
$^{1}$School of Physics and Electronic Engineering, Guangzhou University, Guangzhou 510006, China\\
$^{2}$Shanghai Astronomical Observatory, Chinese Academy of Sciences, 80 Nandan Road, Shanghai 200030, China\\
$^{3}$Key Laboratory for Research in Galaxies and Cosmology, Shanghai Astronomical Observatory, Chinese Academy of Sciences, \\Shanghai 200030, China\\
$^{4}$University of Chinese Academy of Sciences, 19A Yuquanlu, Beijing 100049, China
}

\date{Accepted XXX. Received YYY; in original form ZZZ}

\pubyear{2015}

\begin{document}
\label{firstpage}
\pagerange{\pageref{firstpage}--\pageref{lastpage}}
\maketitle

\begin{abstract}
The outburst and spin-down glitch of magnetars are modeled from the magnetospheric point of view. We try to discuss the following four questions: (1) Which pulsar on the period and peirod-derivative diagram are more likely to show magnetar outburst? (2) Which outburst will make the glitch that triggered the outburst to become a spin-down glitch? (3) Can we model the outburst and spin-down glitch in PSR J1119$-$6127 simultaneously? (4) Why the torque variation is delayed compared with the peak of the X-ray luminosity in 1E 1048.1$-$5937 and PSR J1119$-$6127? It is found that both the global and local twisted magnetic field will affect the radiation and timing behaviors of magnetars. Especially, the delay of torque variations may due to the combined effect of increasing twist in the j-bundle and untwisting of the global magnetosphere. A toy model is build for magnetar outburst and torque variations. It can catch the general trend of magnetar outburst: decaying flux, shrinking hot spot, and torque variations.
\end{abstract}

\begin{keywords}
stars: magnetar -- pulsars: individual (4U 0142+61; PSR J1119$-$6127; 1E 1048.1$-$5937) -- pulsars: general
\end{keywords}


\section{Introduction}

Magnetars may be young and high magnetic field neutron stars (Duncan \& Thompson 1992; Kaspi \& Beloborodov 2017). Compared with normal pulsars, the persistent and burst emission of magnetars may be powered by their magnetic energy. While in normal pulsars, the star's multi-wavelength emissions and associated wind nebula are powered by the star's rotational energy. However, the distinction between normal pulsars and magnetars is not absolute. There may be a continuous distribution from pulsars to magnetars. Possible links between pulsars and magnetars include: (1) Pulsed radio emission of five magnetars are observed up to now (Camilo et al. 2006; Camilo et al. 2007; Levin et al. 2010; Shannon \& Johnston 2013; Esposito et al. 2020; Lower et al. 2020). (2) Several low magnetic field magnetars (with dipole magnetic field order of $\sim 10^{12} \ \rm G$) are discovered (Rea et al. 2010; Zhou et al. 2014). (3) A possible magnetar wind nebula is discovered (Younes et al. 2016). (4) Magnetar activities are seen in one high magnetic field pulsar PSR J1846$-$0258 (Gavriil et al. 2008, which is radio quiet) and one radio-loud high magnetic field pulsar PSR J1119$-$6127 (Archibald et al. 2016; Gogus et al. 2016). It seems that a neutron star with a high magnetic field is more likely to have magnetar activities.
From the energy output point view, which pulsar on the period and period-derivative diagram is more like to show magnetar activities? This is the first question.

During magnetar outburst, various timing event are observed (Dib \& Kaspi 2014). Some of the glitches in magnetars are associated with outburst. While some are not associated outbursts. A special kind of spin-down glitches are observed during magnetar outburst (Archibald et al. 2013). All the spin-down glitches are associated with radiative events (Archibald et al. 2017). During the magnetar outburst of PSR J1119$-$6127, the original glitch finally turned out to be a spin-down glitch (Archibald 2016; Dai et al. 2018; Archibald et al. 2018). Similar things also happened in the outburst from the previous high magnetic field pulsar PSR J1846$-$0258 (Livingstone et al. 2010).  If a glitch triggers an outburst in one magnetar, what's the requirement on the outburst in order to make the glitch to become a spin-down glitch? This is the second question.

During the magnetar outburst of PSR J1119$-$6127, its X-ray flux decrease with time, and its hot spot shrinks with time (Archibald et al. 2018). The glitch at the start finally turned out be a spin-down glitch (Dai et al. 2018; Archibald et al. 2018). Can the outburst (flux decay and shrinking hot spot) and spin-down glitch in PSR J1119$-$6127 be understood simultaneously? This is the third question.

The spin-down glitch reflects the total effects of spin-down torque. By looking carefully at the observations, it is found that the torque variation is always delayed compared with the flux evolution (in 1E 1048.1$-$5927, Archibald et al. 2020; in PSR J1119$-$6127, Archibald et al. 2018). This delay is hard to understand in a simple magnetosphere. Why the torque is delayed compared with the flux evolution? This is the fourth question.

We try to answer the previous four questions from magnetar magnetospheric point of view (Beloborodov 2009; Tong et al. 2013; Tong 2019). In sections two to five, we will answer question one to four, respectively. Especially, in section five, a toy model for magnetar outburst is presented. And possible physics for the delay of torque variation are discussed. Discussion and conclusion are presented in section six.

\section{Which pulsar on the $P$-$\dot{P}$ diagram is more likely to show magnetar outburst?}

\begin{figure}
\centering
\includegraphics[width=0.45\textwidth]{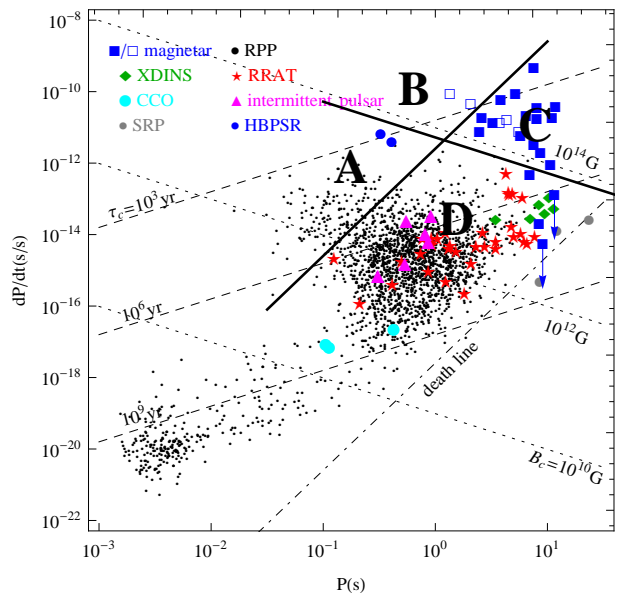}
\caption{Distribution of pulsars and magnetars on the period and period-derivative diagram.
The thick solid line parallel to the death line is the constant rotational energy loss rate line $|\dot{E}_{\rm rot}| = 10^{35} \ \rm erg \ s^{-1}$. The thick solid line parallel to the characteristic magnetic field line is the constant magnetic energy release line $\dot{E}_{\rm p,twist} = 10^{35} \ \rm erg \ s^{-1}$. The $P$-$\dot{P}$ diagram is divided into four regions by these two lines. The discussion about these four regions are presented in the main text. Different symbols represent different kinds of pulsars: black points are normal rotational powered pulsars (Manchester et al. 2005; \url{https://www.atnf.csiro.au/research/pulsar/psrcat/}), red stars are rotating radio transients (RRATs, McLaughlin et al. 2006), pink triangles are intermittent pulsars (Kramer et al. 2006), blue cycles are the two high magnetic field pulsars showing magnetar-like activities (Gavriil et al. 2008; Archiblad et al. 2016), blue squares are magnetars (empty blue squares are radio emitting magnetars, Olausen \& Kaspi 2014; \url{http://www.physics.mcgill.ca/~pulsar/magnetar/main.html}), green diamonds are X-ray dim isolated neutron stars (Kaplan \& van Kerkwijk 2011), cyan cycles are central compact objects (Gotthelf et al. 2013), gray cycles are the three slow radio pulsars with the longest pulsation period (Tan et al. 2018). }
\label{fig_PPdot}
\end{figure}

The X-ray luminosity of persistent magnetars and peak luminosity of transient magnetars is about $10^{35} \ \rm erg \ s^{-1}$ (Coti Zelati et al. 2018). Therefore, lines of constant rotational energy loss rate of $10^{35} \ \rm erg \ s^{-1}$ and lines of constant magnetic energy release rate of $10^{35} \ \rm erg \ s^{-1}$ can be drawn on the $P$-$\dot{P}$ diagram and compared with each other. For a typical moment of inertia of $10^{45} \ \rm g \ cm^2$, the rotational energy loss rate is
\begin{equation}
|\dot{E}_{\rm rot}| = 3.95\times 10^{46} \frac{\dot{P}}{P^3} \ \rm erg \ s^{-1}.
\end{equation}
A constant rotational energy loss rate of $\dot{E}_{\rm rot} =10^{35} \ \rm erg \ s^{-1}$ corresponds to a line parallel to the death line on the $P$-$\dot{P}$ diagram, see figure \ref{fig_PPdot}.

The magnetic energy release rate is more complicated. It may include magnetic energy release from the crustal field (Vigan\`{o} et al. 2013), from local twisted magnetic field in the magnetosphere (Beloborodov 2009), or from open field line regions of a globally twisted magnetosphere (Tong 2019). Taken the globally twisted magnetosphere as an example. For a globally twisted self-similar magnetic field, the radial dependence of the magnetic field is: $B \propto r^{-(2+n)}$ (Wolfson 1995). The magnetic free energy is\footnote{The relation between the parameter $n$ and the maximum twist is: $\Delta \phi_{\rm max} \approx 2\sqrt{35(1-n)/16}$. The parameter $n$ will be used. See Tong (2019) for more discussions.} (Tong 2019)
\begin{equation}\label{eqn_Emf}
E_{\rm mf} = 0.5 (1-n)^{1.5} \frac{1}{12} B_{\rm p}^2 R^3,
\end{equation}
where $B_{\rm p}$ is the surface polar magnetic field in the absence of twist, $R$ is the neutron star radius (taken as $10 \ \rm km$ during the calculations). The magnetic energy release rate and corresponding untwisting timescale depends on the acceleration potential etc (Beleborodov 2009; Tong 2019). We may get an empirical value about the untwisting timescale from magnetar outburst observations (Coti Zelati et al. 2018). The typical flux decay timescale is about one year. This may also be taken as the untwisting timescale of the twisted magnetosphere: $\tau \sim 1 \ \rm yr$. Therefore, the semi-empirical magnetic energy release rate is
\begin{equation}\label{eqn_Edotp}
 \dot{E}_{\rm p,twist} \sim E_{\rm mf}/\tau.
\end{equation}

For typical value of $n=0.5$, the magnetic free energy is $E_{\rm mf} \sim 10^{44} B_{\rm p14}^2 \ \rm erg$, where $B_{\rm p14}$ is the magnetic field in units of $10^{14} \ \rm G$. The corresponding magnetic energy release rate for $n=0.5$ is $\dot{E}_{\rm p,twist} \sim 10^{37} B_{\rm p14}^2 \ \rm erg \ s^{-1}$.
If most of the magnetic energy released is converted to X-ray luminosity (an efficiency about $1$), then the corresponding X-ray luminosity is about $10^{37} B_{\rm p14}^2\ \rm erg \ s^{-1}$. For a typical X-ray luminosity of magnetars about $10^{35} \ \rm erg \ s^{-1}$ (persistent source or peak luminosity during the outburst), the required magnetic field of is about $B_{\rm p14} \sim 0.1$, or $B_{\rm p} \sim 10^{13} \ \rm G$. This is much smaller than the characteristic magnetic field of some magnetars. However, the spin-down torque of magnetars may be enhanced due to the twist of the magnetic field, or an enhanced particle wind (Harding et al. 1999; Tong et al. 2013). The true dipole magnetic field may be only $0.1$ times the characteristic magnetic field which is obtained by assuming pure magnetic dipole braking (Tong et al. 2013). For a typical characteristic dipole field of $10^{14} \ \rm G$, the typical physical magnetic field is about $10^{13} \ \rm G$. This is consistent with the magnetic field inferred from luminosity considerations.

The constant magnetic energy release line $\dot{E}_{\rm p,twist} = 10^{35} \ \rm erg \ s^{-1}$ can also be ploted on the $P$-$\dot{P}$ diagram. According to the above discussions, the true magnetic field is related with the characteristic magnetic field $B_{\rm p} = 0.1 B_{\rm c}$, where $B_{\rm c} =6.4\times 10^{19} \sqrt{P \dot{P}} \ \rm G$ is the characteristic magnetic field. This corresponds to line parallel to the characteristic magnetic field line in the $P$-$\dot{P}$ diagram, see figure \ref{fig_PPdot}.


From Figure \ref{fig_PPdot}, the four regions ABCD have different properties. These four regions may corresponds to different sub-populations of pulsars.
\begin{enumerate}
  \item Region A may correspond to young pulsars. Pulsars in this region have high rotational energy loss rate. However, their magnetic field is relatively low. Even if their magnetic field is twisted, the corresponding magnetic energy release rate is lower their rotational energy loss rate
      $\dot{E}_{\rm p,twist} < 10^{35} {\rm \ erg \ s^{-1}} < \dot{E}_{\rm rot}$. Therefore, emission of pulsars in region A are dominated by their rotational energy loss rate. They are young and energetic pulsars.
  \item Region B may correspond high magnetic field pulsars. Pulsars in this region have both high rotational energy loss rate and high magnetic energy release rate (if their magnetosphere are activated by the twist). Both of the two high magnetic field pulsars are located near region B. Two of the radio emitting magnetars (1E 1547.0$-$5408 and Swift J1818.0$-$1607) are also located on the edge of region B. Furthermore, the radio magnetar Swift J1818.0$-$1607 may be a transition source between normal radio pulsars and the previous four radio emitting magnetars (Lower et al. 2020).
  \item The majority of magnetars are located in region C. Pulsars in this region have low rotational energy loss rate. However, their magnetic field is relatively high. When the pulsar's magnetosphere is twisted, their magnetic energy release rate can be higher than the rotational energy loss rate
      $\dot{E}_{\rm p,twist} > 10^{35} {\rm \  erg \ s^{-1}} > \dot{E}_{\rm rot}$. Pulsars in this region can be persistent magnetars or transient magnetars.
  \item Pulsars in region D have low rotational energy loss rate $\dot{E}_{\rm rot} < 10^{35} \ \rm erg \ s^{-1}$.
  At the same time, their magnetic field is also relatively low. Even if their magnetosphere are twisted, the corresponding magnetic energy release rate is also relatively low $\dot{E}_{\rm p,twist} < 10^{35} \ \rm erg \ s^{-1}$. Therefore, pulsars in region D are hard to detect using high energy observations. However, pulsars can also have pulsed radio emissions. Therefore, pulsars in region D should be mainly detected via their radio emission. Most of the normal pulsars lies in this region.
\end{enumerate}

The definition of ABCD regions are consistent with the general discussions of high magentic field/low magnetic field, high rotational energy loss rate/low rotational energy loss rate. What we have done is: we define a magnetic energy release rate and compare it with the rotational energy loss rate directly (i.e. they have the same dimension). The magnetic energy release rate is determined by the magnetic field (equation (\ref{eqn_Emf}) and (\ref{eqn_Edotp})). It is natural that it is consistent with the discussion of high magnetic field/low magnetic field neutron stars. The boundary between the four regions are not definite. It is only a crude boundary between the relative strength of rotational energy output and magnetic energy output. There should be a smooth transition between different regions. Furthermore, when pulsars in region A are evolved in age, they may all go to region D. Pulsars in region B and C will also go to region D considering evolution of the magnetospheric physics or magnetic field decay (Kou et al. 2019).

Pulsars in region C and D can be both detected via their radio emissions. 
There are five radio emitting magnetars at present. However, their radio properties are very different from normal radio pulsar in region D. From this aspect we may infer that the radio emission of the radio emitting magnetars may originate from their magnetic energy. The magnetar's radio emission have different properties because they originate from a different energy reservoir. However, in principle, magnetars can have rotation powered radio emissions with properties similar to that of normal pulsars (Zhang 2002). The recently discovered radio magnetar Swift J1818.0$-$1607 may represent a transiton object between normal radio pulsars and the previous four radio emitting magnetars (Esposito et al. 2020; Lower et al. 2020).

The above definition of the magnetic energy release line is based on the global twisted magnetosphere. If the magnetosphere is locally twisted, or the magnetic energy release in the crust dominates, the total magnetic free is also proportional to $E_{\rm mf} \propto B_{\rm p}^2 \propto P \dot{P}$. Assuming an empirical constant magnetic energy release timescale, the typical luminosity during outburst is about $L_{\rm x} \propto E_{\rm mf}/\tau \propto P \dot{P}$. It will also be a line parallel to the characteristic magnetic field line. The differences may be only quantitative. Therefore, the discussions about the four regions on the $P$-$\dot{P}$ diagram is still valid. This treat will also be used in the following in building a toy model for magnetar outburst.

\section{Which outburst will make the glitch to become a spin-down glitch?}

If a glitch triggers an outburst in a magnetar, the particle outflow during the outburst will also be enhanced. This will result in an enhanced spin-down rate of the magnetar (Harding et al. 1999; Tong et al. 2013). More rotational energy will be carried away by the outflowing particles. The net effect of the rotational evolution of the magnetar may be a spin-down glitch (Tong 2014), contrary to normal glitches in pulsars and some magnetars. During a glitch, the increase of the rotational energy of the neutron star is
\begin{equation}
\Delta E_{\rm rot} = 2 E_{\rm rot} \frac{\Delta \nu}{\nu}.
\end{equation}
The typical glitch size in magnetars is about $\frac{\Delta \nu}{\nu} \sim 10^{-6}$ (Dib \& Kaspi 2014).
During the outburst, the rotational energy loss rate of the magnetar is enhanced due to an enhanced particle wind (Tong et al. 2013)
\begin{equation}
\dot{E}_{\rm w} = \dot{E}_{\rm d} \left( \frac{L_{\rm p}}{\dot{E}_{\rm d}} \right)^{1/2},
\end{equation}
where $\dot{E}_{\rm w}$ is the rotational energy loss rate due to the enhanced particle wind, $\dot{E}_{\rm d} = 2\mu^2 \Omega^4/3c^3$ is rotational energy loss rate in the case of magnetic dipole braking, $L_{\rm p}$ is the particle wind luminosity. For magnetars, the particle wind luminosity may be comparable to the X-ray luminosity and higher than its rotational energy loss rate: $L_{\rm p} \sim L_{\rm x} > \dot{E}_{\rm rot}$ (Thompson \& Duancan 1996; Duncan 2000; Harding et al. 1999; Tong et al. 2013).
The typical observational duration of magnetar spin-down glitch may be about tens of days (Archibald et al. 2017). The particle luminosity may be taken as constant during this time interval. The total rotational energy carried away by the enhanced particle wind for a time interval of $\Delta t$ is
\begin{equation}
  \Delta E_{\rm w} =\dot{E}_{\rm w} \Delta t.
\end{equation}

In order to make the glitch to become a spin-down glitch, the rotational energy carried away by the enhanced particle wind should be larger than the increase due to the glitch at the start
\begin{equation}
  \Delta E_{\rm w} \ge \Delta E_{\rm rot}.
\end{equation}
Or more explicitly
\begin{equation}
  \dot{E}_{\rm d} \left( \frac{L_{\rm p}}{\dot{E}_{\rm d}} \right)^{1/2} \Delta t \ge 2 E_{\rm rot} \frac{\Delta \nu}{\nu}.
\end{equation}
For an over recovery with a $Q$ factor about ten (Archibald et al. 2017), more rotational energy should be carried away
\begin{equation}
  \dot{E}_{\rm d} \left( \frac{L_{\rm p}}{\dot{E}_{\rm d}} \right)^{1/2} \Delta t \approx Q \times 2 E_{\rm rot} \frac{\Delta \nu}{\nu}.
\end{equation}
After some manipulation, the final result is (numerical factor concering 2/3 are dropped out)
\begin{equation}
  L_{\rm p}^{1/2} \Delta t \approx \frac{c^{3/2} I}{\mu} Q \frac{\Delta \nu}{\nu}.
\end{equation}
This equation means that for a glitch in a magnetar ($\frac{\Delta \nu}{\nu}$), if it finally turned out to be spin-down glitch with a $Q$ factor, the corresponding particle luminosity and duration of the enhanced particle wind should be meet the above requirement.

For typical parameters of moment of inertial $I \sim 10^{45} \ \rm g \ cm^{2}$, dipole magnetic moment
$\mu \sim 4.4\times 10^{31} \ \rm G \ cm^3$ (for typical magnetic field about the quantum critical field $4.4\times 10^{13} \ \rm G$), glitch size $\frac{\Delta \nu}{\nu} \sim 10^{-6}$, an over recovery $Q$ factor $Q \sim 10$, and particle luminosity similar to the X-ray luminosity $L_{\rm p} \sim L_{\rm x} \sim 10^{35} \ \rm erg \ s^{-1}$, the required duration of particle wind $\Delta t$ is about 40 days
\begin{eqnarray}
\nonumber
  \Delta t &\approx& 40 \left( \frac{\Delta\nu /\nu}{10^{-6}} \right) \left( \frac{Q}{10} \right)
  \left( \frac{B}{4.4\times 10^{13} \ \rm G} \right)^{-1}\\
  &&\left( \frac{L_{\rm p}}{10^{35} \ \rm erg \ s^{-1}} \right)^{-1/2}\ \rm day.
\end{eqnarray}
If the Q factor is about one, the required duration of the particle wind $\Delta t$ is about several days.
This means that a larger over recovery Q factor will require a longer duration of the particle wind. This is in general consistent with results summaried in Table 3 in Archibald et al. (2017). This is especially true for the two spin-down glitches in 4U 0142+61: larger Q factor, longer duration of the spin-down timescale. Note that the observations of magnetars, and thus any timing information associated with them, is heavily biased to surrounding times of outbursts. All the spin-down events presented in Archibald et al. (2017) are associated with outbursts. On the other hand, very few magnetars (most notably the five bright magnetars in Dib \& Kaspi (2014)) are monitored outside times of bright outbursts. That is we are unlikely to find ``radition-quiet'' spin down events outside of the well monitored subset (but see Younes et al. 2020; Tong 2020).
Future more spin-down glitches will help to test the above constraints.

\section{Combination: outburst and spin-down glitch in PSR J1119$-$6127}

There is a peak luminosity during magnetar outburst (Coti Zelati et al. 2018). Therefore, low luminosity magnetars (or transient magnetars) are more likely to show a large dynamical range during the outburst. The radio-loud high magnetic field pulsar PSR J1119$-$6127 is similar to transient magnetars in this aspect (Archibald et al. 2018). The magnetar XTE J1810$-$197 is a typical example of transient magnetars. During its outburst decay phase both a decreasing X-ray luminosity and shrinking hot spot are observed (Alford \& Halpern 2016). However, observationally a neutron star radius of $30 \ \rm km$ is required (Alford \& Halpern 2016). This is unreasonable large and may due to other uncertainties. In the case of PSR J1119$-$6127, this does not happen. Furthermore, during the outburst decay phase of PSR J1119$-$6127 a spin-down glitch is also observed (Dai et al. 2018; Archibald et al. 2018). This will make the magnetar outburst in PSR J1119$-$6127 very interesting.

In the twisted local magnetic field model for magnetar outburst (Beloborodov 2009), the X-ray luminosity is due to magnetic energy release during the untwisting process of the magnetosphere. In the globally twisted magnetosphere model, the magnetar may have large polar caps despite its long period (Tong 2019). Particle outflow in the open field line regions may also result in the untwisting of the magnetosphere (Tong 2019)
\begin{equation}
  \frac{E_{\rm mf}}{dt} = -\dot{E}_{\rm p,twist}.
\end{equation}
In the above discussions, estimations on the magnetic energy release rate are given. Physically, the particle luminosity is determined by the outflowing current and acceleration potential
\begin{equation}\label{eqn_untwisting}
  \dot{E}_{\rm p,twist} = 2 I_{\rm pc} \Delta V,
\end{equation}
where $I_{\rm pc}$ is the polar cap current (can be modeled as the Goldreich-Julian current, Tong 2019), $\Delta V$ is the acceleration potential for each particle, a factor of $2$ considers there two polar caps. In Tong (2019), the maximum acceleration potential is used. This is in order to make a unified description of magnetars and pulsars. For PSR J1119$-$6127, its pulsation period is about $0.4 \ \rm s$ (Weltevrede et al. 2011). It is shorter than that of magnetars. Therefore, the corresponding maximum acceleration potential will be very large. However, as in the case of pulsars, there may a critical acceleration potential in the magnetosphere (Medin \& Lai 2010; Tong \& Xu 2012). The physical acceleration potential can not be much higher than this critical value because the gap will be ceased by the cascade when the critical acceleration potential is reached. Therefore, for PSR J1119$-$6127 a constant acceleration potential is chosen $\Delta V \sim (10^{12} -10^{13}) \ \rm V$ (Tong \& Xu 2012; Kou \& Tong 2015). The acceleration potential in the polar acceleration gap depends weakly on the rotation of neutron star (Ruderman \& Sutherland 1975). It can be approximated by the constant acceleration potential.

\begin{figure}
\centering
\begin{minipage}{0.45\textwidth}
 \includegraphics[width=0.95\textwidth]{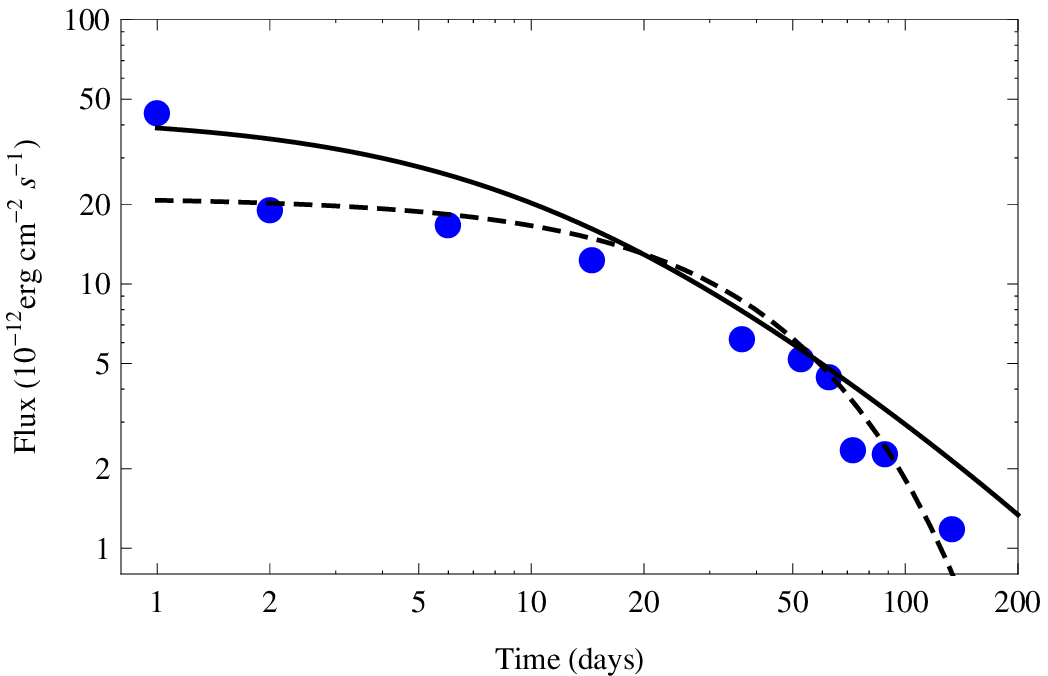}
\end{minipage}
\begin{minipage}{0.45\textwidth}
 \includegraphics[width=0.95\textwidth]{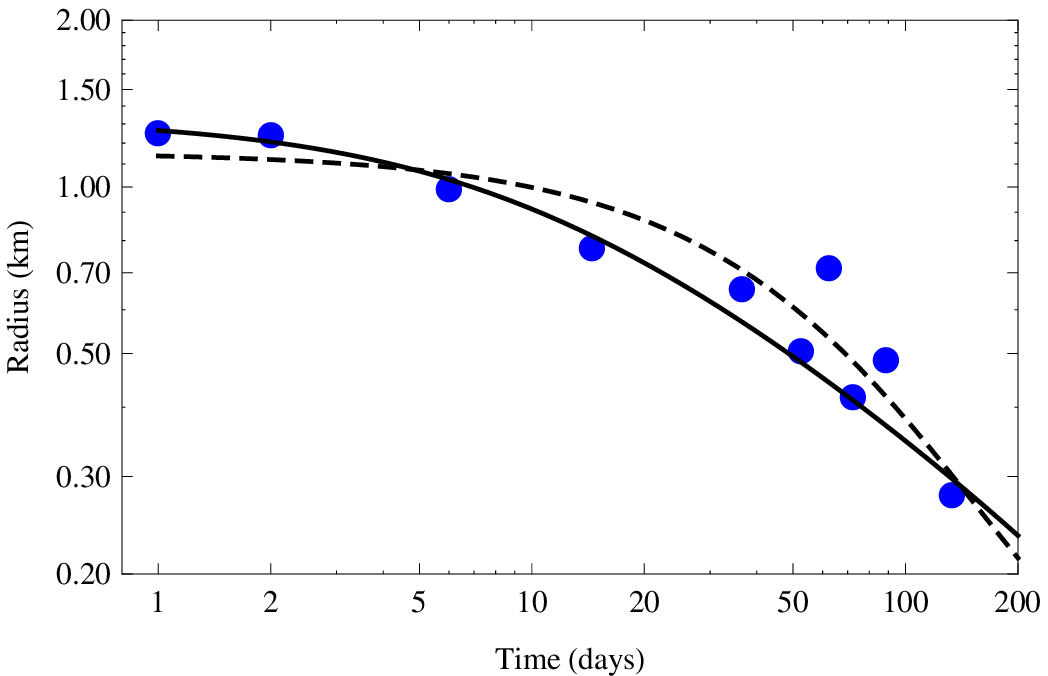}
\end{minipage}
\caption{X-ray flux and hot spot radius of PSR J1119$-$6127. The blue points are observations (table 2 in Archibald et al. 2018). The time of the first observation is taken as day the first. The black lines are the model calculations in the globally twisted magnetosphere. The dashed lines are calculations using the toy model in section 5.}
\label{fig_FluxRadius}
\end{figure}

The magnetosphere untwisting equation (\ref{eqn_untwisting}) is an ordinary differential equation for the parameter $n$. When $n(t)$ is known, the corresponding magnetic free energy and its time derivative is known. And the time derivative derivative of the magnetic free energy $\dot{E}_{\rm mf}$ corresponds to the X-ray luminosity of magnetars, by assuming an energy conversion efficiency of one. The polar cap of the neutron star is related to the parameter $n$ (Tong 2019)
\begin{equation}\label{eqn_thetapc}
  \sin\theta_{\rm pc} = \sqrt{\frac{(R/R_{\rm lc})^n}{(15+17n)/32}},
\end{equation}
where $R$ is the neutron star radius, and $R_{\rm lc}$ is the light cylinder radius.
Therefore, the shrinking hot spot (which is approximately: $R\sin\theta_{\rm pc}$) can also be model by the untwisting globally twisted magnetosphere. The initial value of the hot spot radius determines the initial value of the parameter $n$. The peak luminosity during the outburst determines the magnetic field. The only left free parameter in solving equation (\ref{eqn_untwisting}) is the acceleration potential. For PSR J1119$-$6127, it is found that for a constant acceleration potential of $\Delta V=(1-2)\times 10^{13} \ \rm V$ can explains both the flux and hot spot as a function of time. This value of acceleration potential is similar to that of normal pulsars (Medin \& Lai 2010; Tong \& Xu 2012). For $n(0)=0.4$, a surface polar magnetic field in the absence of twist of $B_{\rm p} =10^{13} \ \rm G$, and $\Delta V=2\times 10^{13} \ \rm V$, the flux and hot spot for PSR J1119$-$6127 is shown in figure \ref{fig_FluxRadius}. If only one polar cap is seen, and for an energy conversion efficiency about one, the theoretical X-ray luminosity should be $\frac12$ of the total magnetic energy release rate.  Observationally, a spherical blackbody emitting surface is assumed. Therefore, the observational reported hot spot radius is actually $\frac12$ of the polar cap radius. These two factors have been taken into consideration in figure \ref{fig_FluxRadius}. The model calculations in the globally twisted magnetosphere can catch the general trend of outburst in PSR J1119$-$6127.

Even at the peak outburst, the X-ray luminosity of PSR J1119$-$6127 is still less than its rotational energy loss rate (Archibald et al. 2018). This means that the above discussion about the origin of spin-down glitch does not apply in PSR J1119$-$6127. There may be two possibilities for the enhanced torque and spin-down glitch in PSR J1119$-$6127.
\begin{enumerate}
  \item A stronger particle wind. If the energy conversion efficiency from particle luminosity to X-ray luminosity is much smaller than one (as in the case of normal pulsars), then the actual particle luminosity can be higher than the rotational energy loss rate. The above discussion in Section 3 about wind enhanced torque can be applied to PSR J1119$-$6127.
  \item Enhanced torque due to an enhanced magnetic field. Both the magnetic dipole braking and the wind braking model of magnetars assumes a large scale dipolar field. When the large scale magnetic field is a twisted dipole magnetic field, it is possible that the torque is also enhanced compared with the magnetic dipole braking case. During the untwisting process, the torque will decay with time. The net effect of the enhanced spin-down torque may make the glitch to become a spin-down glitch.
\end{enumerate}

Both of the above two scenarios will result in positive correlation between the X-ray luminosity and the spin-down torque. However, observationally, the torque variation is delayed compared with the evolution of the X-ray luminosity (Dai et al. 2018; Archibald et al. 2018). The delay of torque variations have been observed many times in 1E 1048.1$-$5937 (Archibald et al. 2020). This may tell us that the magnetosphere of magnetars should be more complicated than the simple self-similar solution.

\section{A toy model for magnetar outburst and delay of torque variations}

\subsection{A toy model for magnetar outburst}

During magnetar outburst, a decaying flux, shrink hot spot, decaying torque, spin-down glitch, softening spectra etc are often seen (Kaspi \& Beloborodov 2017; Coti Zelati et al. 2018).
The internal origin may have difficulties in explaining the shrink hot spot (Vigan\`o et al. 2013). Detailed modeling of the magnetosphere may require semi-analytical or numerical calculations (Beloborodov et al. 2009; Tong 2019). If we relax some of the assumptions, simple analytical formula may be obtained for magnetar outburst. This kind of toy model can catch the general trend of magnetar outburst (decaying flux, shrinking hot spot etc). It may be a bridge magnetar theory and observations.

Irrespective of the location of the twist (local or global etc), the magnetic free energy is decaying during the outburst. By introducing some ``relaxation time'' $\tau$, the evolution of the magnetic free energy is
\begin{equation}
  \frac{d E_{\rm mf}}{d t} = - \frac{E_{\rm mf}}{\tau}.
\end{equation}
For a constant relaxation time, the magnetic free energy decays with time exponentially:
\begin{equation}
  E_{\rm mf}(t) = E_{\rm mf,0} e^{-t/\tau},
\end{equation}
where $E_{\rm mf,0}$ is the initial magnetic free energy. This will result in an exponentially decaying magnetic energy release rate
\begin{equation}
  |\dot{E}_{\rm mf}| = \frac{E_{\rm mf,0}}{\tau} e^{-t/\tau}.
\end{equation}
This may corresponds to the exponential flux decay in observations: $L_{\rm x}(t) = L_{\rm x,0} e^{-t/\tau}$.

In the globally twisted  magnetar magnetosphere model, the evolution of the magnetic free energy corresponds to the evolution of the parameter $n$, see equation (\ref{eqn_Emf}). In equation (\ref{eqn_Emf}), the magnetic free energy depends on the parameter $n$ is a nonlinear way. It can be approximated in a linear way
\begin{equation}
  0.5 (1-n) \frac{1}{12} B_{\rm p}^2 R^3 = E_{\rm mf}.
\end{equation}
By solving this equation, the parameter $n$ evolves with time as
\begin{equation}\label{eqn_n(t)}
  n(t) = 1-(1-n_0) e^{-t/\tau},
\end{equation}
where $n_0$ is the initial parameter of $n$. It is related to the initial magnetic free energy and magnetic field strength. However, it can also be determined directly from the observations. The magnetar polar cap is related to the parameter $n$ through equation (\ref{eqn_thetapc}). The polar cap radius may corresponds to the observed hot spot radius. From the initial observation of the hot spot radius, the initial parameter $n_0$ can be determined. The evolution of the hot spot radius can be modeled by combining equation (\ref{eqn_thetapc}) and (\ref{eqn_n(t)}). Furthermore, equation (\ref{eqn_thetapc}) can also be further simplified by $n\approx 1$
\begin{equation}
  \theta_{\rm pc} = \left( \frac{R}{R_{\rm lc}} \right)^{n/2}.
\end{equation}
And the corresponding polar cap radius is
\begin{eqnarray}\label{eqn_Rpc}
  R_{\rm pc} &=& R \left( \frac{R}{R_{\rm lc}} \right)^{n/2} \\
  &=& R \left( \frac{\Omega R}{c} \right)^{n/2},
\end{eqnarray}
where $n$ is given by equation (\ref{eqn_n(t)}).

For high magnetic field pulsar, their angular velocity $\Omega$ is relative large (compared with magnetars). Therefore, the parameter $n$ may be near 1 in order to explain the hot spot observations (e.g., a hot spot of $1 \ \rm km$). This means that the magnetosphere of high magnetic field pulsars does not deviates from the dipole case significantly. For magnetars with a small anuglar velocity $\Omega$, in order to explain the hot spot observations the parameter $n$ should generally smaller than the high magnetic field pulsar case. This means that the magnetosphere of magnetars deviates from the dipole case significantly during outburst. Observationally, the radio and X-ray pulse profile of magnetar may be more complex and may evolve with time significantly during outburst. This is in general consistent with magnetar observations.

\begin{figure}
\centering
\begin{minipage}{0.45\textwidth}
 \includegraphics[width=0.95\textwidth]{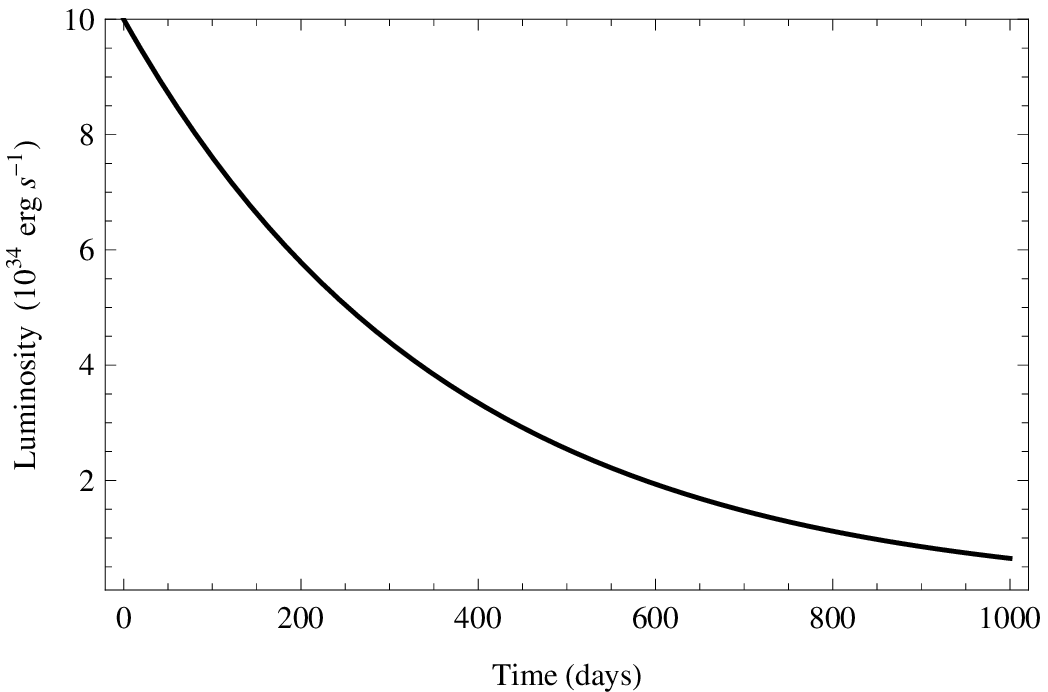}
\end{minipage}
\begin{minipage}{0.45\textwidth}
 \includegraphics[width=0.95\textwidth]{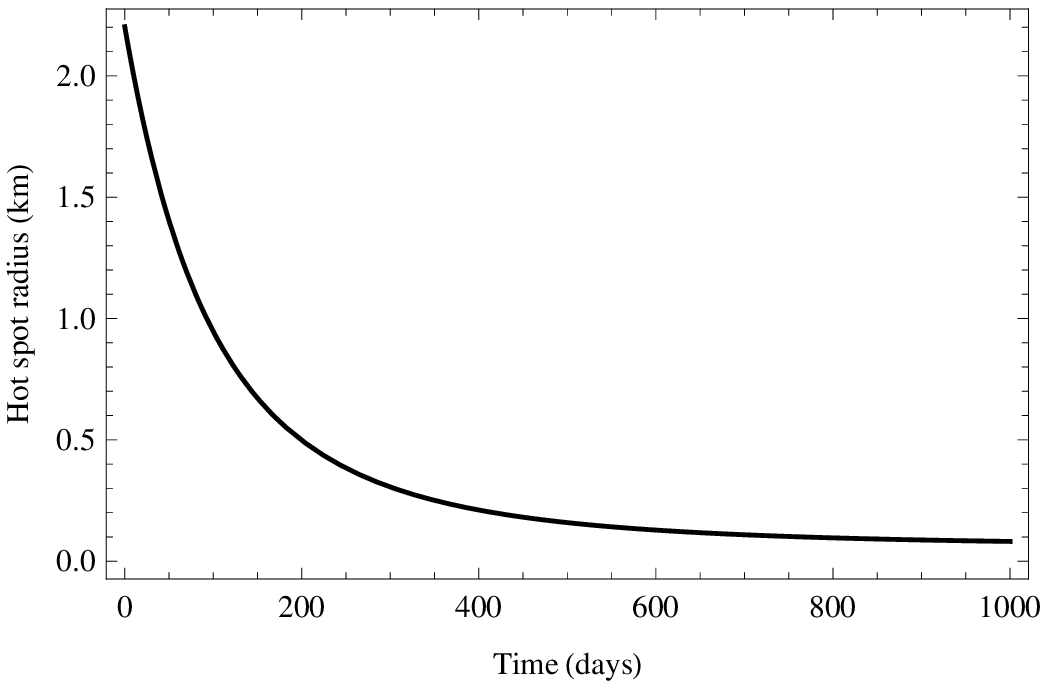}
\end{minipage}
\begin{minipage}{0.45\textwidth}
 \includegraphics[width=0.95\textwidth]{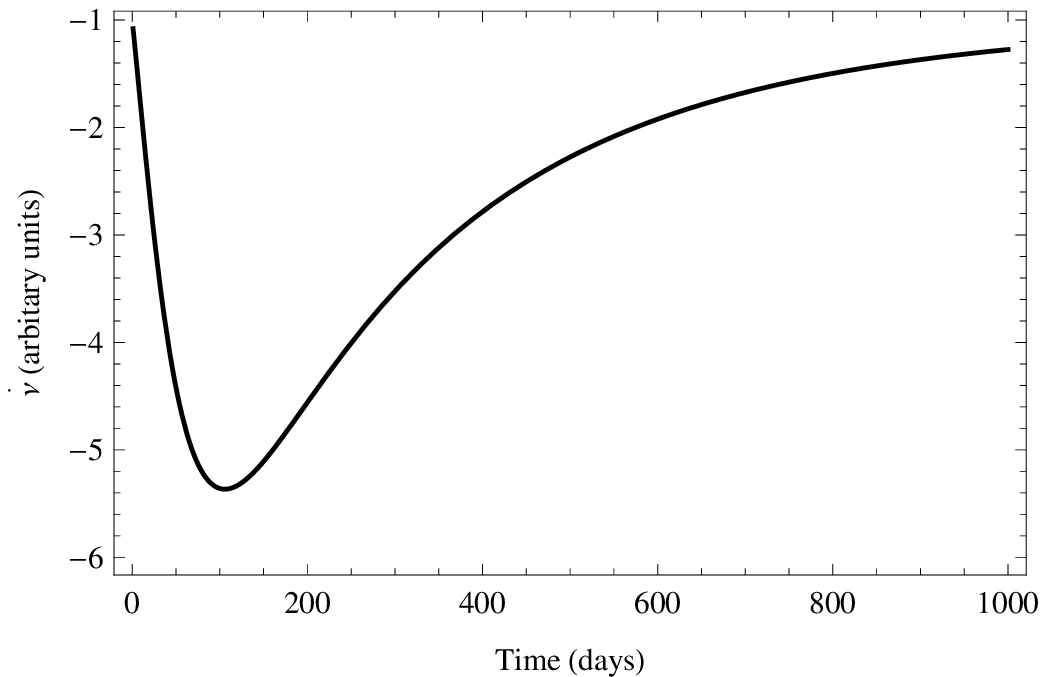}
\end{minipage}
\caption{Luminosity and hot spot evolution of magnetars in the toy model. The bottom panel demonstrates the possible result of torque delay compared with the flux.}
\label{fig_toy_model}
\end{figure}

For typical parameters of $E_{\rm mf,0} \sim 3\times 10^{42} \ \rm erg$, and $\tau \sim 1 \ \rm yr$, the correspond initial X-ray luminosity is about $10^{35} \ \rm erg \ s^{-1}$. For an initial parameter of $n_0 \sim 0.5$, the hot spot radius is about $1 \ \rm km$. Using these typical parameters, the luminosity and hot radius evolution is shown is figure \ref{fig_toy_model}. The toy model for flux decay and shrinking hot spot can also be applied to the previous observations of PSR J1119-6127. The result of toy model is shown in figure \ref{fig_FluxRadius}. For the shrinking hot spot, the best fit parameters are $n_0=0.39$, and $\tau=154$ days. The value of $n_0$ is consistent with the magnetospheric modeling. For the flux decay, however, a different decay time scale is needed  $\tau=40$ days. Observationally, the flux decay will require more than one exponential decaying component (Scholz et al. 2014; Coti Zelati et al. 2018; Archibald et al. 2020). Our one exponential decaying model in Section 5.1 may be too simplified for flux decay of PSR J1119-6127.

\subsection{Delay of torque variations}

The spin-down glitch is due to the integration effect of the enhanced torque. The observations also tell us that the torque variation is delayed compared with the peak of the flux (Dai et al. 2018; Archibald et al. 2018). This have happened many times in the case of 1E 1048.1$-$5937 (Archibald et al. 2020). Some guess can be make about the delay of torque variations based on previous models.

For a twisted magnetosphere of magnetars, it is natural that there are twist in both the closed and open field line regions (Glampedakis et al. 2014). In order to model the true situation, there are two extremes: (1) Consider the twisted magnetosphere as a globally twisted self-similar magnetosphere (Thompson et al. 2002; Pavan et al. 2009; Tong 2019). (2) Consider some dominate local twist (Beloborodov 2009; Huang et al. 2016). In the global twisted magnetosphere, due to the inflation of the magnetic field lines in the radial direction, the magnetar can have large polar caps (Tong 2019). During the untwisting process, a decreasing X-ray luminosity and a shrinking hot spot is expected. In the twisted local magnetic field scenario (Beloborodov 2009), the j-bundle may be taken as the effective polar cap. Therefore, a decreasing flux and shrinking hot spot is also expected during outburst. Observationally, two exponential decaying component is found in 1E 1048.1$-$5937 (Archibald et al. 2020). This may corresponds to the two contributing sources in the magnetosphere: the component with short/long timescale due to  untwisting of the local/global component, respectively.

However, during untwisting process the j-bundle will have an increasing twist. This will make the magnetic field in the j-bundle increase with time, until it saturates. The saturation is reached when the toroidal field is comparable to the poloidal field (Uzdensky 2002). If the j-bundle is coincide with the polar cap regions, it  may make the effective magnetic field in the polar cap region increase with time: $B(t)/B_0 \propto (1-e^{-t/\tau_1})$. At the same time, untwisting of the global magnetosphere will make the twist decay with time, therefore $B(t)/B_0 \propto e^{-t/\tau_2}$. By combining these two factors, the effective magnetic field may evolve with time in the following form
\begin{equation}\label{eqn_B(t)}
  \frac{B(t)}{B_0} = 1+ A (1-e^{-t/\tau_1}) e^{-t/\tau_2},
\end{equation}
where $B_0$ is the initial untwisted magnetic field, $A$ is amplitude factor, $\tau_1$ and $\tau_2$ are the untwisting timescale for the local twist and global twist, respectively. The spin-down torque is expected to be proportional to $B(t)^2$. Therefore, we can calculate the variation of the torque compared with the untwisted case (or persistent case). The result is shown in the bottom of figure \ref{fig_toy_model}. Typical parameters are: an amplitude of $A=2$, $\tau_1$ taken as $50$ days (the shorter timescale for the flux decay, Archibal et al. 2020).  $\tau_2$ is set to be $\tau_2 =1$ yr, the same as the flux decaying timescale in the previous toy model (which is similar to the longer timescale for the flux decay in 1E 1048.1$-$5937, Archibald et al. 2020). The toy model and the discussion here can catch the general behavior of magnetar outburst: decaying flux, shrinking hot spot, torque variations and its delay.

From equation (\ref{eqn_B(t)}), for $t=0$, $B(t)/B_0 =1$, and for $t\sim \tau_1 \ll \tau_2$, $B(t)/B(0) \sim 1+A$. The amplitude factor $A$ represents the enhancement of magnetic field, e.g. at the light cylinder. The neutron star rotational energy loss rate is proportional to $B_{\rm lc}^2 c R_{\rm lc}^2$ (Goldreich \& Julian 1969).
Therefore, the rotational energy loss rate and torque will be enhanced by a factor of $(1+A)^2$. An upper limit can be obtained on the
amplitude factor $A$. For a globally twisted magnetosphere, the magnetic field will decrease with radius in the form of $r^{-(2+n)}$ (Wolfson 1995). For the magnetic dipole case, $n=1$. For a hot spot about 1 km, the parameter $n$ is about $n=0.5$. The magnetic field at the light cylinder will be enhanced by about $(R_{\rm lc}/R_{\rm ns})^{1/2} \sim 100$, for typical magnetar rotational period. The rotational energy loss rate and toque will be amplified by about $10^4$. This strong enhancement may happen during magnetar giant flares. An estimation of the amplitude factor $A$ can be done in the wind braking model. For a strong particle wind $L_{\rm p}\gg \dot{E}_{\rm rot}$, the magnetic field will be opened at $r_{\rm open} \ll R_{\rm lc}$ (Harding et al. 1999; Tong et al. 2013). The new radius $r_{\rm open}$ will play the role the light cylinder radius in the case of wind braking. The final rotational energy loss rate will be amplified by a factor $(L_{\rm p}/\dot{E}_{\rm d})^{1/2}$ (Harding et al. 1999; Tong et al. 2013). Therefore, the magnetic field at the light cylinder radius will be amplified by $(L_{\rm p}/\dot{E}_{\rm d})^{1/4} \sim (L_{\rm x} / |\dot{E}_{\rm rot}|)^{1/4}$. For typical X-ray luminosity about $L_{\rm x} \sim 10^{35} \ \rm erg \ s^{-1}$ and quiescent rotational energy loss rate $10^{31}-10^{33} \ \rm erg \ s^{-1}$, the magnetic field will be amplified by a factor about $(3-10)$. Therefore, the amplitude factor $A$ will be in the range about $(2-10)$. This may be the typical range for the amplitude factor $A$.

The enhancement of the spin-down torque and delay of torque variations may take different forms in different outbursts. If the outburst is dominated by crustal magnetic energy release or local magnetic energy release in the closed field line regions of the magnetosphere, then the outburst may have very little torque variations (Scholz et al. 2014; see also Tong \& Xu 2013 for an alternative.). A shrinking hot spot may indicate a magnetospheric origin, and vice versa. Torque variation will be present if (1) the local twist affect not only the closed field line region but also the open field line region (Beloborodov 2009), or (2) the magnetosphere can be considered as globally twisted (i.e. a large portion of the magnetosphere are twisted, Thompson et al. 2002; Pavan et al. 2009; Glampedakis et al. 2014; Tong 2019). During the outburst decay phase, the torque are general expected to decay with time (e.g., XTE J1810-197, Levin et al. 2019). The delay of torque variation will occur when both (1) the j-bundle of the twisted local field lies in the open field line region, (2) the magnetosphere is also twisted globally. The competition of these two factors will result in a delayed torque variation, as that in equation (\ref{eqn_B(t)}).

\section{Discussion and conclusion}

Some general discussions are presented for magnetar outburst, spin-down glitches, and delay of torque variations. The observations of outburst in PSR J1119$-$6127 and 1E 1048.1$-$5937 are discussed in more detail. The typical example of magnetar outburst in XTE J1810$-$197 have been modeled in previous works (Beloborodov 2009; Tong 2019). It is straightforward to do similar calculations for other sources: evolution of their flux, hot spot, torques etc. The model parameters may change from source to source. But the general trend is similar (Coti Zelati et al. 2018).

The coupled evolution of magnetic field and thermal heat may also explain the flux decay of magnetar outburst (Vigan\`o et al. 2013). But it may have difficulty in explaining the shrinking hot spot. The magnetospheric model (both local and global modeling, Beloborodov 2009; Tong 2019) can explain the decreasing flux and shrinking hot spot simultaneously. In explaining the spin-down glitch and torque variations, there may be two case of: (1) the toque variation is dominated by a strong particle wind ($L_{\rm p} \sim L_x > \dot{E}_{\rm rot}$ Tong et al. 2013; and Section 3), (2) the torque variation is due to changes in the magnetic field strength (the corresponding particle wind is not very strong, Section 5.2). In the case of a strong particle wind, the torque variation is directly related to the flux evolution. There will be no delay of torque variations. Some magnetar outburst may corresponds to this case (XTE J1810$-$197, Levin et al. 2019). Both the untwisting global magnetosphere and the local magnetic field may contribution the evolution of the magnetic field. This case may result in the delay of torque variations. However, the evolution of twist (not self-similar) in both the open and closed field line regions may be beyond the reach of analytical modeling. Numerical calculations in this direction are need.

Some suggestion for the observers when discussing magnetar outburst.
\begin{enumerate}
  \item If they want to discussion the decaying flux and shrink hot spot, they can use the toy model presented in Section 5.1. The initial flux and decaying timescale $\tau$ can be obtained by fitting the flux evolution. The initial parameter of $n_0$ can be obtained by the initial observations of hot spot. The evolution of hot spot can be model using equation (\ref{eqn_Rpc}). After using this toy model, more detailed model calculations can be done (Beloborodov 2009; Vigan\`o et al. 2013; Tong 2019). If two exponential components are presented in the flux observations, the component with longer timescale should be chosen.
  \item If they want to discuss the spin-down glitch, they can use the constrains presented in Section 3. It is for the strong particle wind case $L_{\rm p} \sim L_{\rm x} > \dot{E}_{\rm rot}$. In the case of $L_{\rm p} \sim L_{\rm x}\le \dot{E}_{\rm rot}$, the corresponding calculations can only be taken as order of magnitude estimations. The torque form based on equation (\ref{eqn_B(t)}) (the torque is $\propto B(t)^2$) can also be employed to discuss the spin-down glitch. Order of magnitude estimation is: the spin-down rate will be amplified by about $A^2$ for a duration about $\tau_1$, for a spin-down glitch with over-recovery factor of $Q$ the final requirement is:
      \begin{equation}
      A^2 \tau_1 \sim Q \tau_{c} \Delta \nu/\nu,
      \end{equation}
      wehere $\tau_{\rm c}$ is the magnetar's characteristic age. Using typical parameters for 4U 0142+61 (Archibald et al. 2017), the amplitude factor $A$ is estimated to be order of unity.
  \item If they want to discuss the delay of torque variations, they can use the calculations in Section 5.2. The flux decay should be model by two exponential component. Then $\tau_1$ and $\tau_2$ can be obtained. The only free parameter is the amplitude factor $A$.
\end{enumerate}

In conclusion: (1) Pulsars with stronger magnetic field are more likely to show magnetar outburst. Because when their magnetosphere are twisted, the resulting magnetic energy release rate is very high (can be higher than their rotational energy loss rate). (2) The particle flux and the outburst duration should meet some requirement in order to make the glitch (which trigger the magnetar outburst) to become a spin-down glitch. (3) Magnetospheric modeling can catch the general trend of outburst in PSR J1119$-$6127. In order to explain its spin-down glitch, some additional inputs may be needed. (4) A toy model for magnetar outburst is build. The delay of the torque variation may be due to combined effects of increasing twist in the j-bundle (which is due to untwisting of some local twist) and a global untwisting magnetosphere.

\section*{Acknowledgements}

We would like to thank the referee very much for helpful comments.
H.Tong is supported by NSFC (11773008).
L.H. thanks for the support by NSFC (11933007), Research Program of Fundamental and Frontier Sciences, CAS (ZDBS-LY-SLH011), and Key Research Program of Frontier Sciences, CAS (QYZDJ-SSW-SLH057).

\section*{Data availability}
The data underlying this article are available in the article. 






\bsp	
\label{lastpage}
\end{document}